**RESEARCH ARTICLE**  Open Access

# Estimating the sample mean and standard deviation from the sample size, median, range and/or interquartile range

Xiang Wan[1†], Wenqian Wang[2†], Jiming Liu[1] and Tiejun Tong[3*]

**Abstract**

**Background:** In systematic reviews and meta-analysis, researchers often pool the results of the sample mean and standard deviation from a set of similar clinical trials. A number of the trials, however, reported the study using the median, the minimum and maximum values, and/or the first and third quartiles. Hence, in order to combine results, one may have to estimate the sample mean and standard deviation for such trials.

**Methods:** In this paper, we propose to improve the existing literature in several directions. First, we show that the sample standard deviation estimation in Hozo et al.'s method (BMC Med Res Methodol 5:13, 2005) has some serious limitations and is always less satisfactory in practice. Inspired by this, we propose a new estimation method by incorporating the sample size. Second, we systematically study the sample mean and standard deviation estimation problem under several other interesting settings where the interquartile range is also available for the trials.

**Results:** We demonstrate the performance of the proposed methods through simulation studies for the three frequently encountered scenarios, respectively. For the first two scenarios, our method greatly improves existing methods and provides a nearly unbiased estimate of the true sample standard deviation for normal data and a slightly biased estimate for skewed data. For the third scenario, our method still performs very well for both normal data and skewed data. Furthermore, we compare the estimators of the sample mean and standard deviation under all three scenarios and present some suggestions on which scenario is preferred in real-world applications.

**Conclusions:** In this paper, we discuss different approximation methods in the estimation of the sample mean and standard deviation and propose some new estimation methods to improve the existing literature. We conclude our work with a summary table (an Excel spread sheet including all formulas) that serves as a comprehensive guidance for performing meta-analysis in different situations.

## Background

In medical research, it is common to find that several similar trials are conducted to verify the clinical effectiveness of a certain treatment. While individual trial study could fail to show a statistically significant treatment effect, systematic reviews and meta-analysis of combined results might reveal the potential benefits of treatment. For instance, Antman et al. [1] pointed out that systematic reviews and meta-analysis of randomized control trials would have led to earlier recognition of the benefits of thrombolytic therapy for myocardial infarction and may save a large number of patients.

Prior to the 1990s, the traditional approach to combining results from multiple trials is to conduct narrative (unsystematic) reviews, which are mainly based on the experience and subjectivity of experts in the area [2]. However, this approach suffers from many critical flaws. The major one is due to inconsistent criteria of different reviewers. To claim a treatment effect, different reviewers may use different thresholds, which often lead to opposite

*Correspondence: tongt@hkbu.edu.hk
†Equal Contributors
[3]Department of Mathematics, Hong Kong Baptist University, Kowloon Tong, Hong Kong
Full list of author information is available at the end of the article conclusions from the same study. Hence, from the mid-1980s, systematic reviews and meta-analysis have become an imperative tool in medical effectiveness measurement. Systematic reviews use specific and explicit criteria to identify and assemble related studies and usually provide a quantitative (statistic) estimate of aggregate effect over all the included studies. The methodology in systematic reviews is usually referred to as meta-analysis. With the combination of several studies and more data taken into consideration in systematic reviews, the accuracy of estimations will get improved and more precise interpretations towards the treatment effect can be achieved via meta-analysis.

In meta-analysis of continuous outcomes, the sample size, mean, and standard deviation are required from included studies. This, however, can be difficult because results from different studies are often presented in different and non-consistent forms. Specifically in medical research, instead of reporting the sample mean and standard deviation of the trials, some trial studies only report the median, the minimum and maximum values, and/or the first and third quartiles. Therefore, we need to estimate the sample mean and standard deviation from these quantities so that we can pool results in a consistent format. Hozo et al. [3] were the first to address this estimation problem. They proposed a simple method for estimating the sample mean and the sample variance (or equivalently the sample standard deviation) from the median, range, and the size of the sample. Their method is now widely accepted in the literature of systematic reviews and meta-analysis. For instance, a search of Google Scholar on November 12, 2014 showed that the article of Hozo et al.'s method has been cited 722 times where 426 citations are made recently in 2013 and 2014.

In this paper, we will show that the estimation of the sample standard deviation in Hozo et al.'s method has some serious limitations. In particular, their estimator did not incorporate the information of the sample size and so consequently, it is always less satisfactory in practice. Inspired by this, we propose a new estimation method that will greatly improve their method. In addition, we will investigate the estimation problem under several other interesting settings where the first and third quartiles are also available for the trials.

Throughout the paper, we define the following summary statistics:

$a = $ the minimum value,
$q_1 = $ the first quartile,
$m = $ the median,
$q_3 = $ the third quartile,
$b = $ the maximum value,
$n = $ the sample size.

The $\{a, q_1, m, q_3, b\}$ is often referred to as the 5-number summary [4]. Note that the 5-number summary may not always be given in full. The three frequently encountered scenarios are:

$$\mathcal{C}_1 = \{a, m, b; n\},$$
$$\mathcal{C}_2 = \{a, q_1, m, q_3, b; n\},$$
$$\mathcal{C}_3 = \{q_1, m, q_3; n\}.$$

Hozo et al.'s method only addressed the estimation of the sample mean and variance under Scenario $\mathcal{C}_1$ while Scenarios $\mathcal{C}_2$ and $\mathcal{C}_3$ are also common in systematic review and meta-analysis. In Sections 'Methods' and 'Results', we study the estimation problem under these three scenarios, respectively. Simulation studies are conducted in each scenario to demonstrate the superiority of the proposed methods. We conclude the paper in Section 'Discussion' with some discussions and a summary table to provide a comprehensive guidance for performing meta-analysis in different situations.

## Methods
### Estimating $\bar{X}$ and $S$ from $\mathcal{C}_1$

Scenario $\mathcal{C}_1$ assumes that the median, the minimum, the maximum and the sample size are given for a clinical trial study. This is the same assumption as made in Hozo et al.'s method. To estimate the sample mean and standard deviation, we first review the Hozo et al.'s method and point out some limitations of their method in estimating the sample standard deviation. We then propose to improve their estimation by incorporating the information of the sample size.

Throughout the paper, we let $X_1, X_2, \ldots, X_n$ be a random sample of size $n$ from the normal distribution $N(\mu, \sigma^2)$, and $X_{(1)} \leq X_{(2)} \leq \cdots \leq X_{(n)}$ be the ordered statistics of $X_1, X_2, \cdots, X_n$. Also for the sake of simplicity, we assume that $n = 4Q + 1$ with $Q$ being a positive integer. Then

$$\begin{aligned}
a = X_{(1)} &\leq X_{(2)} \leq \cdots \leq X_{(Q+1)} = q_1 \\
&\leq X_{(Q+2)} \leq \cdots \leq X_{(2Q+1)} = m \\
&\leq X_{(2Q+2)} \leq \cdots \leq X_{(3Q+1)} = q_3 \\
&\leq X_{(3Q+2)} \leq \cdots \leq X_{(4Q+1)} = X_{(n)} = b.
\end{aligned} \quad (1)$$

In this section, we are interested in estimating the sample mean $\bar{X} = \sum_{i=1}^{n} X_i$ and the sample standard deviation $S = \left[\sum_{i=1}^{n}(X_i - \bar{X})^2/(n-1)\right]^{1/2}$, given that $a, m, b,$ and $n$ of the data are known.

*Hozo et al.'s method*
For ease of notation, let $M = 2Q + 1$. Then, $M = (n + 1)/2$. To estimate the mean value, Hozo et al. applied the following inequalities:

$$a \leq X_{(1)} \leq a$$
$$a \leq X_{(i)} \leq m \quad (i = 2, \ldots, M - 1)$$
$$m \leq X_{(M)} \leq m$$
$$m \leq X_{(i)} \leq b \quad (i = M + 1, \ldots, n - 1)$$
$$b \leq X_{(n)} \leq b.$$

Adding up all above inequalities and dividing by $n$, we have $LB_1 \leq \bar{X} \leq UB_1$, where the lower and upper bounds are

$$LB_1 = \frac{a + m}{2} + \frac{2b - a - m}{2n},$$
$$UB_1 = \frac{m + b}{2} + \frac{2a - m - b}{2n}.$$

Hozo et al. then estimated the sample mean by

$$\frac{LB_1 + UB_1}{2} = \frac{a + 2m + b}{4} + \frac{a - 2m + b}{4n}. \quad (2)$$

Note that the second term in (2) is negligible when the sample size is large. A simplified mean estimation is given as

$$\bar{X} \approx \frac{a + 2m + b}{4}. \quad (3)$$

For estimating the sample standard deviation, by assuming that the data are non-negative, Hozo et al. applied the following inequalities:

$$aX_{(1)} \leq X_{(1)}^2 \leq aX_{(1)}$$
$$aX_{(i)} \leq X_{(i)}^2 \leq mX_{(i)} \quad (i = 2, \ldots, M - 1)$$
$$mX_{(M)} \leq X_{(M)}^2 \leq mX_{(M)} \quad (4)$$
$$mX_{(i)} \leq X_{(i)}^2 \leq bX_{(i)} \quad (i = M + 1, \ldots, n - 1)$$
$$bX_{(n)} \leq X_{(n)}^2 \leq bX_{(n)}.$$

With some simple algebra and approximations on the formula (4), we have $LSB_1 \leq \sum_{i=1}^{n} X_i^2 \leq USB_1$, where the lower and upper bounds are

$$LSB_1 = a^2 + m^2 + b^2 + (M - 2)\frac{a^2 + am + m^2 + mb}{2},$$
$$USB_1 = a^2 + m^2 + b^2 + (M - 2)\frac{am + m^2 + mb + b^2}{2}.$$

Then by (3) and the approximation $\sum_{i=1}^{n} X_i^2 \approx (LSB_1 + USB_1)/2$, the sample standard deviation is estimated by $S = \sqrt{S^2}$, where

$$S^2 = \frac{1}{n - 1}\left(\sum_{i=1}^{n} X_i^2 - n\bar{X}^2\right)$$
$$\approx \frac{1}{n - 1}\left(a^2 + m^2 + b^2 + \frac{(n - 3)}{2}\frac{(a + m)^2 + (m + b)^2}{4} - \frac{n(a + 2m + b)^2}{16}\right).$$

When $n$ is large, it results in the following well-known range rule of thumb:

$$S \approx \frac{b - a}{4}. \quad (5)$$

Note that the range rule of thumb (5) is independent of the sample size. It may not work well in practice, especially when $n$ is extremely small or large. To overcome this problem, Hozo et al. proposed the following improved range rule of thumb with respect to the different size of the sample:

$$S \approx \begin{cases} \frac{1}{\sqrt{12}}\left[(b - a)^2 + \frac{(a - 2m + b)^2}{4}\right]^{1/2} & n \leq 15 \\ \frac{b - a}{4} & 15 < n \leq 70 \\ \frac{b - a}{6} & n > 70, \end{cases} \quad (6)$$

where the formula for $n \leq 15$ is derived under the equidistantly spaced data assumption, and the formula for $n > 70$ is suggested by the Chebyshev's inequality [5]. Note also that when the data are symmetric, we have $a + b \approx 2m$ and so

$$\frac{1}{\sqrt{12}}\left[(b - a)^2 + \frac{(a - 2m + b)^2}{4}\right]^{1/2} \approx \frac{b - a}{\sqrt{12}}.$$

Hozo et al. showed that the adaptive formula (6) performs better than the original formula (5) in most settings.

*Improved estimation of S*
We think, however, that the adaptive formula (6) may still be less accurate for practical use. First, the threshold values 15 and 70 are suggested somewhat arbitrarily. Second, given the normal data $N(\mu, \sigma^2)$ with $\sigma > 0$ being a finite value, we know that $\sigma \approx (b - a)/6 \to \infty$ as $n \to \infty$. This contradicts to the assumption that $\sigma$ is a finite value. Third, the non-negative data assumption in Hozo et al.'s method is also quite restrictive.

In this section, we propose a new estimator to further improve (6) and, in addition, we remove the non-negative assumption on the data. Let $Z_1, \ldots, Z_n$ be independent and identically distributed (i.i.d.) random variables from the standard normal distribution $N(0, 1)$, and $Z_{(1)} \leq \cdots \leq Z_{(n)}$ be the ordered statistics of $Z_1, \ldots, Z_n$. Then $X_i = \mu + \sigma Z_i$ and $X_{(i)} = \mu + \sigma Z_{(i)}$ for $i = 1, \ldots, n$. In particular, we have $a = \mu + \sigma Z_{(1)}$ and $b = \mu + \sigma Z_{(n)}$. Since $E(Z_{(1)}) = -E(Z_{(n)})$, we have $E(b - a) = 2\sigma E(Z_{(n)})$. Hence, by letting $\xi(n) = 2E(Z_{(n)})$, we choose the following estimation for the sample standard deviation:

$$S \approx \frac{b-a}{\xi(n)}. \qquad (7)$$

Note that $\xi(n)$ plays an important role in the sample standard deviation estimation. If we let $\xi(n) \equiv 4$, then (7) reduces to the original rule of thumb in (5). If we let $\xi(n) = \sqrt{12}$ for $n \leq 15$, 4 for $15 < n \leq 70$, or 6 for $n > 70$, then (7) reduces to the improved rule of thumb (6).

Next, we present a method to approximate $\xi(n)$ and establish an adaptive rule of thumb for standard deviation estimation. By David and Nagaraja's method [6], the expected value of $Z_{(n)}$ is

$$E(Z_{(n)}) = n \int_{-\infty}^{\infty} z [\Phi(z)]^{n-1} \phi(z) dz,$$

where $\phi(z) = \frac{1}{\sqrt{2\pi}} e^{-z^2/2}$ is the probability density function and $\Phi(z) = \int_{-\infty}^{z} \phi(t) dt$ is the cumulative distribution function of the standard normal distribution. For ease of reference, we have computed the values of $\xi(n)$ by numerical integration using the computer in Table 1 for $n$ up to 50. From Table 1, it is evident that the adaptive formula (6) in Hozo et al.'s method is less accurate and also less flexible.

When $n$ is large (say $n > 50$), we can apply Blom's method [7] to approximate $E(Z_{(n)})$. Specifically, Blom suggested the following approximation for the expected values of the order statistics:

$$E(Z_{(r)}) \approx \Phi^{-1}\left(\frac{r-\alpha}{n-2\alpha+1}\right), \quad r = 1, \ldots, n, \qquad (8)$$

where $\Phi^{-1}(z)$ is the inverse function of $\Phi(z)$, or equivalently, the upper $z$th percentile of the standard normal distribution. Blom observed that the value of $\alpha$ increases as $n$ increases, with the lowest value being 0.330 for $n = 2$. Overall, Blom suggested $\alpha = 0.375$ as a compromise value for practical use. Further discussion on the choice of $\alpha$ can be seen, for example, in [8] and [9]. Finally, by (7) and (8) with $r = n$ and $\alpha = 0.375$, we estimate the sample standard deviation by

$$S \approx \frac{b-a}{2\Phi^{-1}\left(\frac{n-0.375}{n+0.25}\right)}. \qquad (9)$$

In the statistical software R, the upper $z$th percentile $\Phi^{-1}(z)$ can be computed by the command "qnorm(z)".

### Estimating $\bar{X}$ and $S$ from $\mathcal{C}_2$

Scenario $\mathcal{C}_2$ assumes that the first quartile, $q_1$, and the third quartile, $q_3$, are also available in addition to $\mathcal{C}_1$. In this setting, Bland's method [10] extended Hozo et al.'s results by incorporating the additional information of the interquartile range (IQR). He further claimed that the new estimators for the sample mean and standard deviation are superior to those in Hozo et al.'s method. In this section, we first review the Bland's method and point out some limitations of this method. We then, accordingly, propose to improve this method by incorporating the size of a sample.

#### Bland's method

Noting that $n = 4Q + 1$, we have $Q = (n-1)/4$. To estimate the sample mean, Bland's method considered the following inequalities:

$$\begin{aligned}
a &\leq X_{(1)} \leq a \\
a &\leq X_{(i)} \leq q_1 & (i = 2, \ldots, Q) \\
q_1 &\leq X_{(Q+1)} \leq q_1 \\
q_1 &\leq X_{(i)} \leq m & (i = Q+2, \ldots, 2Q) \\
m &\leq X_{(2Q+1)} \leq m \\
m &\leq X_{(i)} \leq q_3 & (i = 2Q+2, \ldots, 3Q) \\
q_3 &\leq X_{(3Q+1)} \leq q_3 \\
q_3 &\leq X_{(i)} \leq b & (i = 3Q+2, \ldots, n-1) \\
b &\leq X_{(n)} \leq b.
\end{aligned}$$

**Table 1 Values of $\xi(n)$ in the formula (7) and the formula (12) for $n \leq 50$**

| n | $\xi(n)$ | n | $\xi(n)$ | n | $\xi(n)$ | n | $\xi(n)$ | n | $\xi(n)$ |
|---|---|---|---|---|---|---|---|---|---|
| 1 | 0 | 11 | 3.173 | 21 | 3.778 | 31 | 4.113 | 41 | 4.341 |
| 2 | 1.128 | 12 | 3.259 | 22 | 3.819 | 32 | 4.139 | 42 | 4.361 |
| 3 | 1.693 | 13 | 3.336 | 23 | 3.858 | 33 | 4.165 | 43 | 4.379 |
| 4 | 2.059 | 14 | 3.407 | 24 | 3.895 | 34 | 4.189 | 44 | 4.398 |
| 5 | 2.326 | 15 | 3.472 | 25 | 3.931 | 35 | 4.213 | 45 | 4.415 |
| 6 | 2.534 | 16 | 3.532 | 26 | 3.964 | 36 | 4.236 | 46 | 4.433 |
| 7 | 2.704 | 17 | 3.588 | 27 | 3.997 | 37 | 4.259 | 47 | 4.450 |
| 8 | 2.847 | 18 | 3.640 | 28 | 4.027 | 38 | 4.280 | 48 | 4.466 |
| 9 | 2.970 | 19 | 3.689 | 29 | 4.057 | 39 | 4.301 | 49 | 4.482 |
| 10 | 3.078 | 20 | 3.735 | 30 | 4.086 | 40 | 4.322 | 50 | 4.498 |

Adding up all above inequalities and dividing by $n$, it results in $LB_2 \leq \bar{X} \leq UB_2$, where the lower and upper bounds are

$$LB_2 = \frac{a + q_1 + m + q_3}{4} + \frac{4b - a - q_1 - m - q_3}{4n},$$
$$UB_2 = \frac{q_1 + m + q_3 + b}{4} + \frac{4a - q_1 - m - q_3 - b}{4n}.$$

Bland then estimated the sample mean by $(LB_2 + UB_2)/2$. When the sample size is large, by ignoring the negligible second terms in $LB_2$ and $UB_2$, a simplified mean estimation is given as

$$\bar{X} \approx \frac{a + 2q_1 + 2m + 2q_3 + b}{8}. \quad (10)$$

For the sample standard deviation, Bland considered some similar inequalities as in (4). Then with some simple algebra and approximation, it results in $LSB_2 \leq \sum_{i=1}^{n} X_i^2 \leq USB_2$, where the lower and upper bounds are

$$LSB_2 = \frac{1}{8} \left[ (n + 3) \left( a^2 + q_1^2 + m^2 + q_3^2 \right) + 8b^2 \right]$$
$$+ (n - 5)(aq_1 + q_1 m + mq_3 + q_3 b) ],$$
$$USB_2 = \frac{1}{8} \left[ 8a^2 + (n + 3) \left( q_1^2 + m^2 + q_3^2 + b^2 \right) \right]$$
$$+ (n - 5)(aq_1 + q_1 m + mq_3 + q_3 b) ].$$

Next, by the approximation $\sum_{i=1}^{n} X_i^2 \approx (LSB_2 + USB_2)/2$,

$$S^2 \approx \frac{1}{16} \left( a^2 + 2q_1^2 + 2m^2 + 2q_3^2 + b^2 \right)$$
$$+ \frac{1}{8}(aq_1 + q_1 m + mq_3 + q_3 b) - \frac{1}{64}(a + 2q_1 + 2m + 2q_3 + b)^2. \quad (11)$$

Bland's method then took the square root $\sqrt{S^2}$ to estimate the sample standard deviation. Note that the estimator (11) is independent of the sample size $n$. Hence, it may not be sufficient for general use, especially when $n$ is small or large. In the next section, we propose an improved estimation for the sample standard deviation by incorporating the additional information of the sample size.

### Improved estimation of S

Recall that the range $b - a$ was used to estimate the sample standard deviation in Scenario $\mathcal{C}_1$. Now for Scenario $\mathcal{C}_2$, since the IQR $q_3 - q_1$ is also known, another approach is to estimate the sample standard deviation by $(q_3 - q_1)/\eta(n)$, where $\eta(n)$ is a function of $n$. Taking both methods into account, we propose the following combined estimator for the sample standard deviation:

$$S \approx \frac{1}{2} \left( \frac{b - a}{\xi(n)} + \frac{q_3 - q_1}{\eta(n)} \right). \quad (12)$$

Following Section 'Improved estimation of S', we have $\xi(n) = 2E(Z_{(n)})$. Now we look for an expression for $\eta(n)$ so that $(q_3 - q_1)/\eta(n)$ also provides a good estimate of $S$. By (1), we have $q_1 = \mu + \sigma Z_{(Q+1)}$ and $q_3 = \mu + \sigma Z_{(3Q+1)}$. Then, $q_3 - q_1 = \sigma \left( Z_{(3Q+1)} - Z_{(Q+1)} \right)$. Further, by noting that $E\left( Z_{(Q+1)} \right) = -E\left( Z_{(3Q+1)} \right)$, we have $E(q_3 - q_1) = 2\sigma E\left( Z_{(3Q+1)} \right)$. This suggests that

$$\eta(n) = 2E\left( Z_{(3Q+1)} \right).$$

In what follows, we propose a method to compute the value of $\eta(n)$. By [6], the expected value of $Z_{(3Q+1)}$ is

$$E\left( Z_{(3Q+1)} \right) = \frac{(4Q+1)!}{(Q)!\,(3Q)!} \int_{-\infty}^{\infty} z[\Phi(z)]^{3Q}[1-\Phi(z)]^Q \phi(z) dz.$$

In Table 2, we provide the numerical values of $\eta(n) = 2E(Z_{(3Q+1)})$ for $Q \leq 50$ using the statistical software R. When $n$ is large, we suggest to apply the formula (8) to approximate $\eta(n)$. Specifically, noting that $Q = (n-1)/4$, we have $\eta(n) \approx 2\Phi^{-1}((0.75n - 0.125)/(n + 0.25))$ for $r = 3Q + 1$ with $\alpha = 0.375$. Then consequently, for the scenario $\mathcal{C}_2$ we estimate the sample standard deviation by

$$S \approx \frac{b - a}{4\Phi^{-1}\left( \frac{n - 0.375}{n + 0.25} \right)} + \frac{q_3 - q_1}{4\Phi^{-1}\left( \frac{0.75n - 0.125}{n + 0.25} \right)}. \quad (13)$$

**Table 2** Values of $\eta(n)$ in the formula (12) and the formula (15) for $Q \leq 50$, where $n = 4Q + 1$

| Q | $\eta(n)$ | Q | $\eta(n)$ | Q | $\eta(n)$ | Q | $\eta(n)$ | Q | $\eta(n)$ |
|---|---|---|---|---|---|---|---|---|---|
| 1 | 0.990 | 11 | 1.307 | 21 | 1.327 | 31 | 1.334 | 41 | 1.338 |
| 2 | 1.144 | 12 | 1.311 | 22 | 1.328 | 32 | 1.334 | 42 | 1.338 |
| 3 | 1.206 | 13 | 1.313 | 23 | 1.329 | 33 | 1.335 | 43 | 1.338 |
| 4 | 1.239 | 14 | 1.316 | 24 | 1.330 | 34 | 1.335 | 44 | 1.338 |
| 5 | 1.260 | 15 | 1.318 | 25 | 1.330 | 35 | 1.336 | 45 | 1.339 |
| 6 | 1.274 | 16 | 1.320 | 26 | 1.331 | 36 | 1.336 | 46 | 1.339 |
| 7 | 1.284 | 17 | 1.322 | 27 | 1.332 | 37 | 1.336 | 47 | 1.339 |
| 8 | 1.292 | 18 | 1.323 | 28 | 1.332 | 38 | 1.337 | 48 | 1.339 |
| 9 | 1.298 | 19 | 1.324 | 29 | 1.333 | 39 | 1.337 | 49 | 1.339 |
| 10 | 1.303 | 20 | 1.326 | 30 | 1.333 | 40 | 1.337 | 50 | 1.340 |

We note that the formula (13) is more concise than the formula (11). The numerical comparison between the two formulas will be given in the section of simulation study.

**Estimating $\bar{X}$ and $S$ from $\mathcal{C}_3$**

Scenario $\mathcal{C}_3$ is an alternative way to report the study other than Scenarios $\mathcal{C}_1$ and $\mathcal{C}_2$. It reports the first and third quartiles instead of the minimum and maximum values. One main reason to report $\mathcal{C}_3$ is because the IQR is usually less sensitive to outliers compared to the range. For the new scenario, we note that Hozo et al.'s method and Bland's method will no longer be applicable. Particularly, if their ideas are followed, we have the following inequalities:

$$
\begin{aligned}
-\infty &\le X_{(i)} \le q_1 & (i=1,\ldots,Q) \\
q_1 &\le X_{(Q+1)} \le q_1 & \\
q_1 &\le X_{(i)} \le m & (i=Q+2,\ldots,2Q) \\
m &\le X_{(2Q+1)} \le m & \\
m &\le X_{(i)} \le q_3 & (i=2Q+2,\ldots,3Q) \\
q_3 &\le X_{(3Q+1)} \le q_3 & \\
q_3 &\le X_{(i)} \le \infty, & (i=3Q+2,\ldots,n)
\end{aligned}
$$

where the first $Q$ inequalities are unbounded for the lower limit, and the last $Q$ inequalities are unbounded for the upper limit. Now adding up all above inequalities and dividing by $n$, we have $-\infty \le \bar{X} \le \infty$. This shows that the approaches based on the inequalities do not apply to Scenario $\mathcal{C}_3$.

In contrast, the following procedure is commonly adopted in the recent literature including [11,12]: "*If the study provided medians and IQR, we imputed the means and standard deviations as described by Hozo et al. [3]. We calculated the lower and upper ends of the range by multiplying the difference between the median and upper and lower ends of the IQR by 2 and adding or subtracting the product from the median, respectively*". This procedure, however, performs very poorly in our simulations (not shown).

*A quantile method for estimating $\bar{X}$ and $S$*

In this section, we propose a quantile method for estimating the sample mean and the sample standard deviation, respectively. In detail, we first revisit the estimation method in Scenario $\mathcal{C}_2$. By (10), we have

$$\bar{X} \approx \frac{a + 2q_1 + 2m + 2q_3 + b}{8} = \frac{a+b}{8} + \frac{q_1 + m + q_3}{4}.$$

Now for Scenario $\mathcal{C}_3$, $a$ and $b$ are not given. Hence, a reasonable solution is to remove $a$ and $b$ from the estimation and keep the second term. By doing so, we have the estimation form as $\bar{X} \approx (q_1 + m + q_3)/C$, where $C$ is a constant. Finally, noting that $E(q_1 + m + q_3) = 3\mu + \sigma E\left(Z_{(Q+1)} + Z_{2Q+1} + Z_{(3Q+1)}\right) = 3\mu$, we let $C = 3$ and define the estimator of the sample mean as follows:

$$\bar{X} \approx \frac{q_1 + m + q_3}{3}. \tag{14}$$

For the sample standard deviation, following the idea in constructing (12) we propose the following estimation:

$$S \approx \frac{q_3 - q_1}{\eta(n)}, \tag{15}$$

where $\eta(n) = 2E\left(Z_{(3Q+1)}\right)$. As mentioned above that $E(q_3 - q_1) = 2\sigma E\left(Z_{(3Q+1)}\right) = \sigma \eta(n)$, therefore, the estimator (15) provides a good estimate for the sample standard deviation. The numerical values of $\eta(n)$ are given in Table 2 for $Q \le 50$. When $n$ is large, by the approximation $E(Z_{(3Q+1)}) \approx \Phi^{-1}\left((0.75n - 0.125)/(n + 0.25)\right)$, we can also estimate the sample standard deviation by

$$S \approx \frac{q_3 - q_1}{2\Phi^{-1}\left(\frac{0.75n - 0.125}{n + 0.25}\right)}. \tag{16}$$

A similar estimator for estimating the standard deviation from IQR is provided in the Cochrane Handbook [13], which is defined as

$$S \approx \frac{q_3 - q_1}{1.35}. \tag{17}$$

Note that the estimator (17) is also independent of the sample size $n$ and thus may not be sufficient for general use. As we can see from Table 2, the value of $\eta(n)$ in the formula (15) converges to about 1.35 when $n$ is large. Note also that the denominator in formula (16) converges to $2 * \Phi^{-1}(0.75)$ which is 1.34898 as $n$ tends to infinity. When the sample size is small, our method will provide more accurate estimates than the formula (17) for the standard deviation estimation.

**Results**

**Simulation study for $\mathcal{C}_1$**

In this section, we conduct simulation studies to compare the performance of Hozo et al.'s method and our new method for estimating the sample standard deviation. Following Hozo et al.'s settings, we consider five different distributions: the normal distribution with mean $\mu = 50$ and standard deviation $\sigma = 17$, the log-normal distribution with location parameter $\mu = 4$ and scale parameter $\sigma = 0.3$, the beta distribution with shape parameters $\alpha = 9$ and $\beta = 4$, the exponential distribution with rate parameter $\lambda = 10$, and the Weibull distribution with shape parameter $k = 2$ and scale parameter $\lambda = 35$. The graph of each of these distributions with the specified parameters is provided in Additional file 1. In each simulation, we first randomly sample $n$ observations and compute the true sample standard deviation using the whole sample. We then use the median, the minimum and maximum values of the sample to estimate the sample standard deviation by the formulas (6) and (9), respectively. To assess the accuracy of the two estimates, we define the relative error of each method as

$$\text{relative error of } S = \frac{\text{the estimated } S - \text{the true } S}{\text{the true } S}. \quad (18)$$

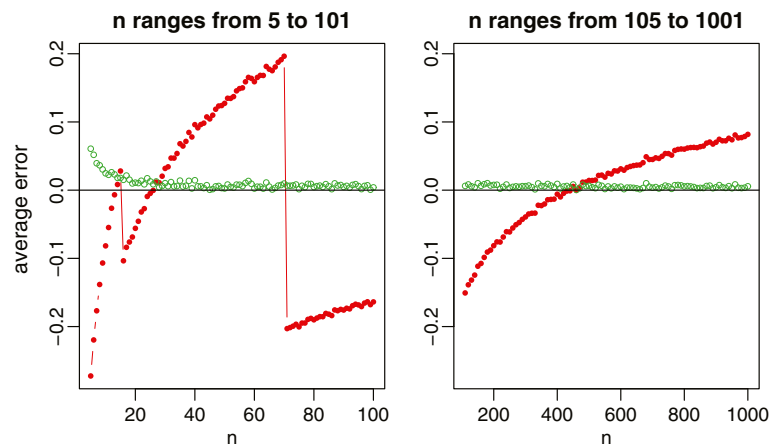

**Figure 1 Relative errors of the sample standard deviation estimation for normal data, where the red lines with solid circles represent Hozo et al.'s method, and the green lines with empty circles represent the new method.**

With 1000 simulations, we report the average relative errors in Figure 1 for the normal distribution with the sample size ranging from 5 to 1001, and in Figure 2 for the four non-normal distributions with the sample size ranging from 5 to 101. For normal data which are most commonly assumed in meta-analysis, our new method provides a nearly unbiased estimate of the true sample standard deviation. Whereas for Hozo et al.'s method, we do observe that the best cutoff value is about $n = 15$ for switching between the estimates $(b-a)/\sqrt{12}$ and $(b-a)/4$, and is about $n = 70$ for switching between $(b-a)/4$ and $(b-a)/6$. However, its overall performance is not satisfactory by noting that the estimate always fluctuates from -20% to 20% of the true sample standard deviation. In addition, we note that $\xi(27) \approx 4$ from Table 1 and $\xi(n) \approx 6$ when $\Phi^{-1}((n-0.375)/(n+0.25)) = 3$, that is, $n = (0.375 + 0.25 * \Phi(3))/(1 - \Phi(3)) \approx 463$. This coincides with the simulation results in Figure 1 where the method $(b-a)/4$ crosses the $x$-axis between $n = 20$ and $n = 30$, and the method $(b-a)/6$ crosses the $x$-axis between $n = 400$ and $n = 500$.

From Figure 2 with the skewed data, our proposed method (9) makes a slightly biased estimate with the relative errors about 5% of the true sample standard deviation. Nevertheless, it is still obvious that the new method is much better compared to Hozo et al.'s method. We also note that, for the beta and Weibull distributions, the best cutoff values of $n$ should be larger than 70 for switching between $(b-a)/4$ and $(b-a)/6$. This again coincides with

Table one in Hozo et al. [3] where the suggested cutoff value is $n = 100$ for Beta and $n = 110$ for Weibull.

**Simulation study for $C_2$**

In this section, we evaluate the performance of the proposed method (13) and compare it to Bland's method (11). Following Bland's settings, we consider (i) the normal distribution with mean $\mu = 5$ and standard deviation $\sigma = 1$, and (ii) the log-normal distribution with location parameter $\mu = 5$ and scale parameter $\sigma = 0.25, 0.5$, and 1, respectively. For simplicity, we consider the sample size being $n = 4Q + 1$, where $Q$ takes values from 1 to 50. As in Section 'Simulation study for $C_1$', we assess the accuracy of the two estimates by the relative error defined in (18).

In each simulation, we draw a total of $n$ observations randomly from the given distribution and compute the true sample standard deviation of the sample. We then use and only use the minimum value, the first quartile, the median, the third quartile, and the maximum value to estimate the sample standard deviation by the formulas (11) and (13), respectively. With 1000 simulations, we report the average relative errors in Figure 3 for the four specified distributions. From Figure 3, we observe that the new method provides a nearly unbiased estimate of the true sample standard deviation. Even for the very highly skewed log-normal data with $\sigma = 1$, the relative error of the new method is also less than 10% for most sample sizes. On the contrary, Bland's method is less satisfactory. As reported in [10], the formula (11) only works for a small range of sample sizes (In our simulations, the range is about from 20 to 40). When the sample size gets larger or the distribution is highly skewed, the sample standard deviations will be highly overestimated. Additionally, we note that the sample standard deviations will be seriously underestimated if $n$ is very small. Overall, it is evident that the new method is better than Bland's method in most settings.

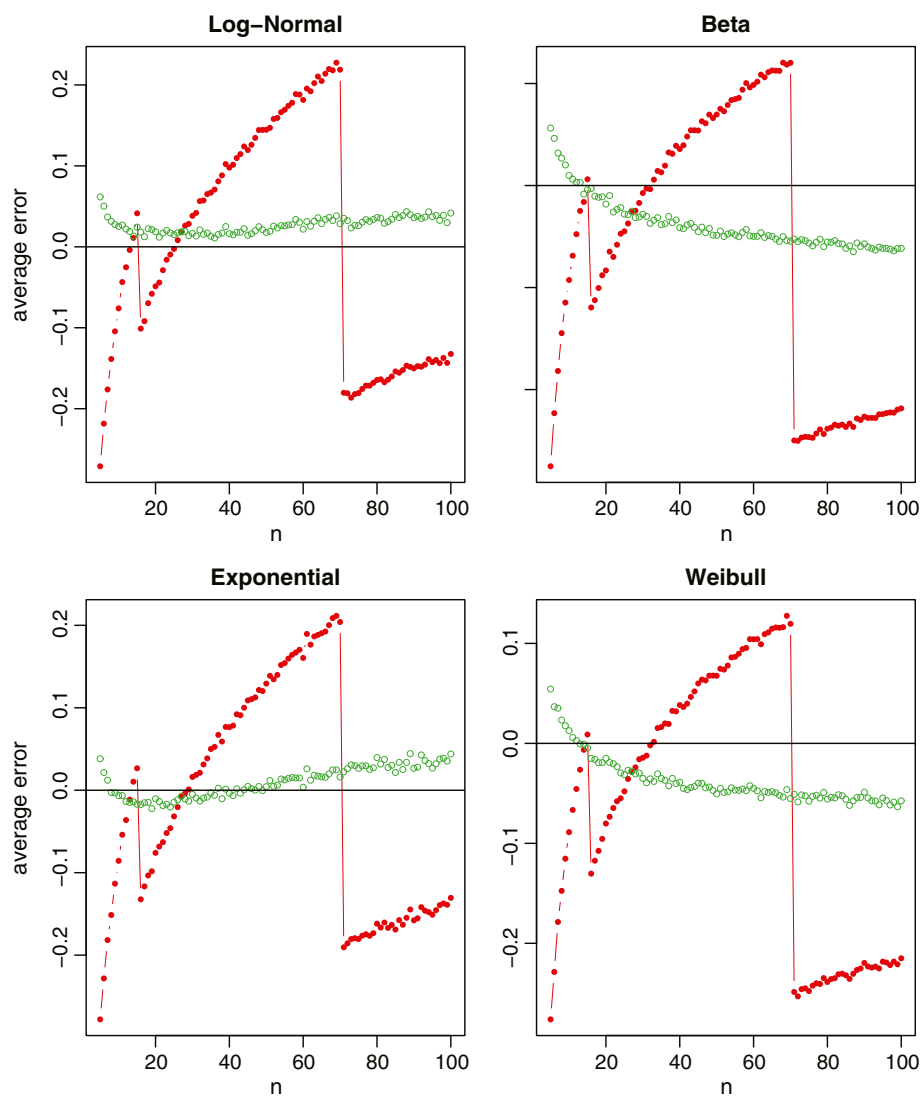

**Figure 2 Relative errors of the sample standard deviation estimation for non-normal data (log-normal, beta, exponential and Weibull), where the red lines with solid circles represent Hozo et al.'s method, and the green lines with empty circles represent the new method.**

**Simulation study for $\mathcal{C}_3$**

In the third simulation study, we conduct a comparison study that not only assesses the accuracy of the proposed method under Scenario $\mathcal{C}_3$, but also addresses a more realistic question in meta-analysis, "*For a clinical trial study, which summary statistics should be preferred to report, $\mathcal{C}_1$, $\mathcal{C}_2$ or $\mathcal{C}_3$? and why?*"

For the sample mean estimation, we consider the formulas (3), (10), and (14) under three different scenarios, respectively. The accuracy of the mean estimation is also assessed by the relative error, which is defined in the same way as that for the sample standard deviation estimation.

Similarly, for the sample standard deviation estimation, we consider the formulas (9), (13), and (15) under three different scenarios, respectively. The distributions we considered are the same as in Section 'Simulation study for $\mathcal{C}_1$', i.e., the normal, log-normal, beta, exponential and Weibull distributions with the same parameters as those in previous two simulation studies.

In each simulation, we first draw a random sample of size $n$ from each distribution. The true sample mean and the true sample standard deviation are computed using the whole sample. The summary statistics are also computed and categorized into Scenarios $\mathcal{C}_1$, $\mathcal{C}_2$ and $\mathcal{C}_3$. We then use the aforementioned formulas to estimate the sample mean and standard deviation, respectively. The sample sizes are $n = 4Q + 1$, where $Q$ takes values from 1 to 50. With 1000 simulations, we report the average relative errors in Figure 4 for both $\bar{X}$ and $S$ with the normal distribution, in Figure 5 for the sample mean estimation with the non-normal distributions, and in Figure 6 for the sample standard deviation estimation with the non-normal distributions.

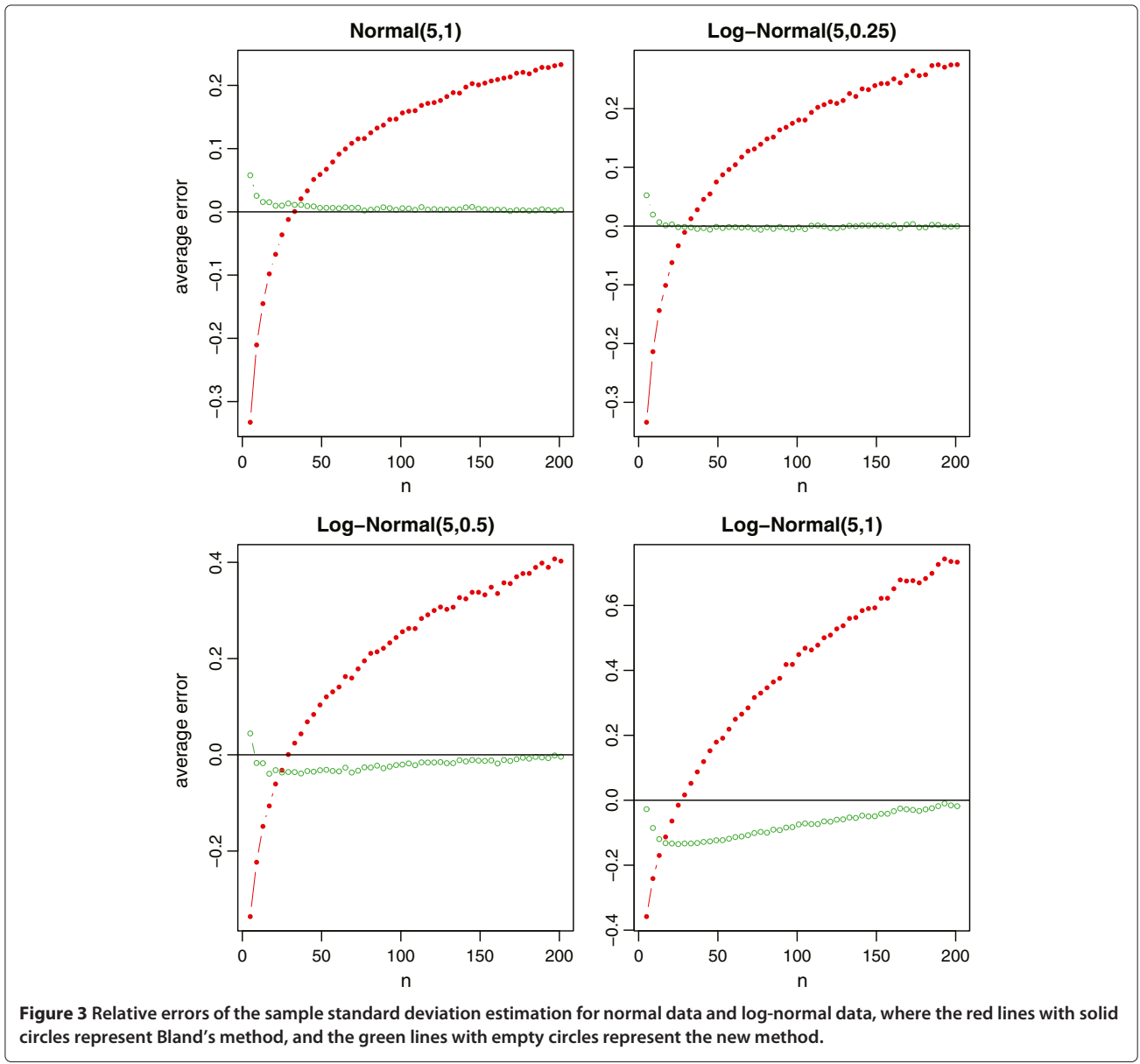

**Figure 3 Relative errors of the sample standard deviation estimation for normal data and log-normal data, where the red lines with solid circles represent Bland's method, and the green lines with empty circles represent the new method.**

For normal data which meta-analysis would commonly assume, all three methods provide a nearly unbiased estimate of the true sample mean. The relative errors in the sample standard deviation estimation are also very small in most settings (within 1% in general). Among the three methods, however, we recommend to estimate $\bar{X}$ and $S$ using the summary statistics in Scenario $\mathcal{C}_3$. One main reason is because the first and third quartiles are usually less sensitive to outliers compared to the minimum and maximum values. Consequently, $\mathcal{C}_3$ produces a more stable estimation than $\mathcal{C}_1$, and also $\mathcal{C}_2$ that is partially affected by the minimum and maximum values.

For non-normal data from Figure 5, we note that the mean estimation from $\mathcal{C}_2$ is always better than that from $\mathcal{C}_1$. That is, if the additional information in the first and third quartiles is available, we should always use such information. On the other hand, the estimation from $\mathcal{C}_2$ may not be consistently better than that from $\mathcal{C}_3$ even though $\mathcal{C}_2$ contains the additional information of minimum and maximum values. The reason is that this additional information may contain extreme values which may not be fully reliable and thus lead to worse estimation.

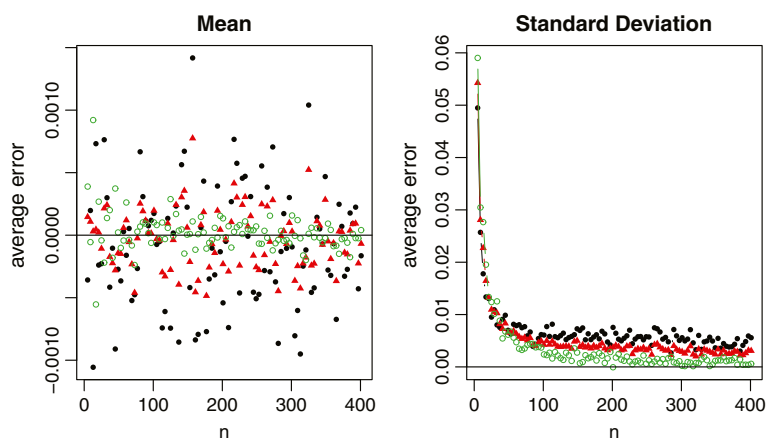

**Figure 4 Relative errors of the sample mean and standard deviation estimations for normal data, where the black solid circles represent the method under scenario $C_1$, the red solid triangles represent the method under scenario $C_2$, and the green empty circles represent the method under scenario $C_3$.**

Therefore, we need to be cautious when making the choice between $C_2$ and $C_3$. It is also noteworthy that (i) the mean estimation from $C_3$ is not sensitive to the sample size, and (ii) $C_1$ and $C_3$ always lead to opposite estimations (one underestimates and the other overestimates the true value). While from Figure 6, we observe that (i) the standard deviation estimation from $C_3$ is quite sensitive to the skewness of the data, (ii) $C_1$ and $C_3$ would also lead to the opposite estimations except for very small sample sizes, and (iii) $C_2$ turns out to be a good compromise for estimating the sample standard deviation. Taking both into account, we recommend to report Scenario $C_2$ in clinical trial studies. However, if we do not have all information in the 5-number summary and have to make a decision between $C_1$ and $C_3$, we recommend $C_1$ for small sample sizes (say $n \leq 30$), and $C_3$ for large sample sizes.

**Discussion**

Researchers often use the sample mean and standard deviation to perform meta-analysis from clinical trials. However, sometimes, the reported results may only include the sample size, median, range and/or IQR. To combine these results in meta-analysis, we need to estimate the sample mean and standard deviation from them. In this paper, we first show the limitations of the existing works and then propose some new estimation methods. Here we summarize all discussed and proposed estimators under different scenarios in Table 3.

We note that the proposed methods are established under the assumption that the data are normally distributed. In meta-analysis, however, the medians and quartiles are often reported when data do not follow a normal distribution. A natural question arises: "*To which extent it makes sense to apply methods that are based on a normal distribution assumption?*" In practice, if the entire sample or a large part of the sample is known, standard methods in statistics can be applied to estimate the skewness or even the density of the population. For the current study, however, the information provided is very limited, say for example, only *a*, *m*, *b* and *n* are given in Scenario 1. Under such situations, it may not be feasible to obtain a reliable estimate for the skewness unless we specify the underlying distribution for the population. Note that the underlying distribution is unlikely to be known in practice. Instead, if we arbitrarily choose a distribution (more likely to be misspecified), then the estimates from the wrong model can be even worse than that from the normal distribution assumption. As a compromise, we expect that the proposed formulas under the normal distribution assumption are among the best we can achieve.

Secondly, we note that even if the means and standard deviations can be satisfyingly estimated from the proposed formulas, it still remains a question to which extent it makes sense to use them in a meta-analysis, if the underlying distribution is very asymmetric and one must assume that they don't represent location and dispersion adequately. Overall, this is a very practical yet challenging question and may warrant more research. In our future research, we propose to develop some test statistics (likelihood ratio test, score test, etc) for pre-testing the hypothesis that the distribution is symmetric (or normal) under the scenarios we considered in this article. The result of the pre-test will then suggest us whether or not we should still include the (very) asymmetric data in the meta-analysis. Other proposals that address this issue will also be considered in our future study.

Finally, to promote the usability, we have provided an Excel spread sheet to include all formulas in Table 3 in Additional file 2. Specifically, in the Excel spread sheet, our proposed methods for estimating the sample mean and standard deviation can be applied by simply inputting the sample size, the median, the minimum and maximum values, and/or the first and third quartiles for the appropriate scenario. Furthermore, for ease of comparison, we have also included Hozo et al.'s method and Bland's method in the Excel spread sheet.

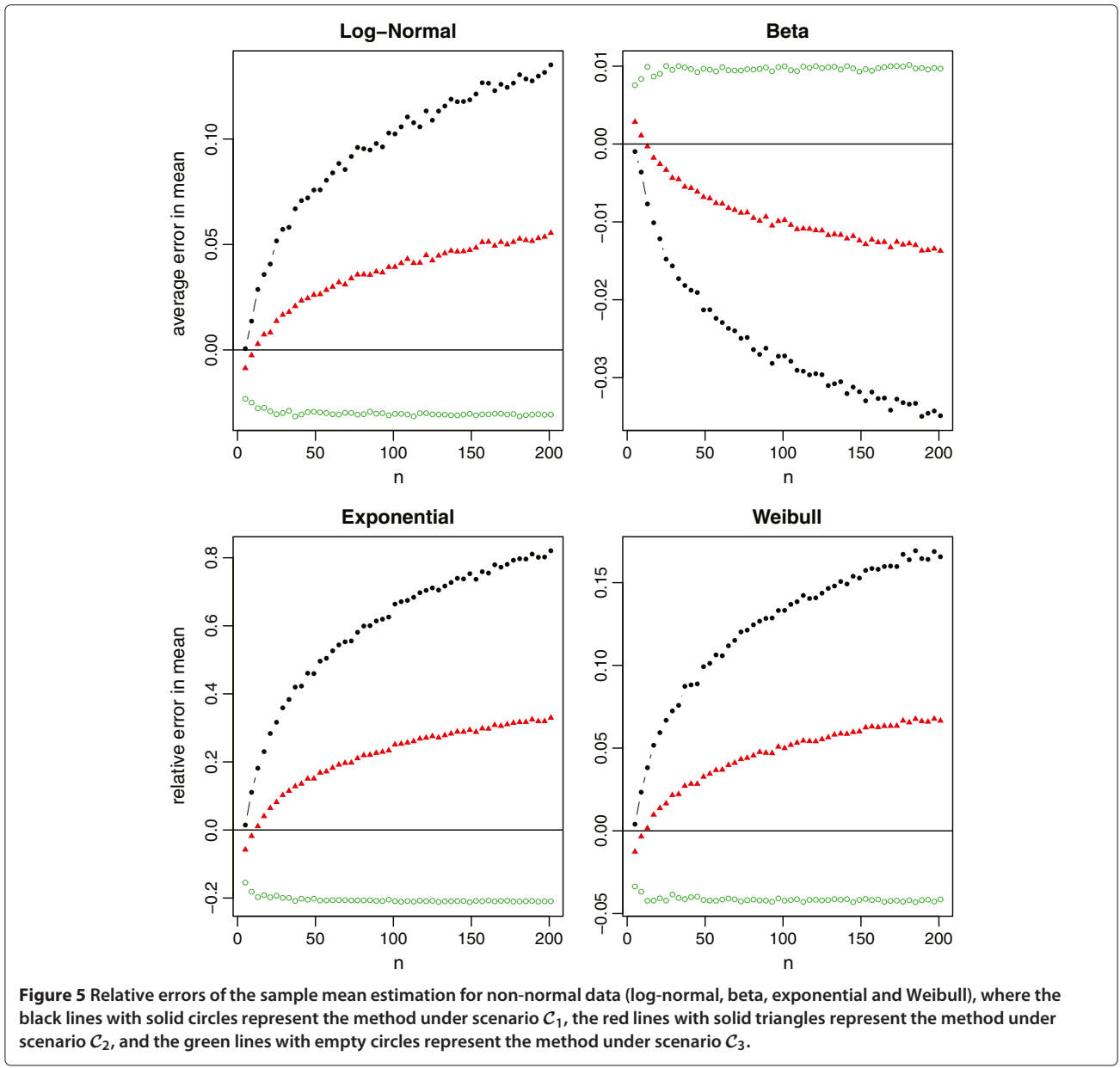

**Figure 5 Relative errors of the sample mean estimation for non-normal data (log-normal, beta, exponential and Weibull), where the black lines with solid circles represent the method under scenario $\mathcal{C}_1$, the red lines with solid triangles represent the method under scenario $\mathcal{C}_2$, and the green lines with empty circles represent the method under scenario $\mathcal{C}_3$.**

## Conclusions

In this paper, we discuss different approximation methods in the estimation of the sample mean and standard deviation and propose some new estimation methods to improve the existing literature. Through simulation studies, we demonstrate that the proposed methods greatly improve the existing methods and enrich the literature.

Specifically, we point out that the widely accepted estimator of standard deviation proposed by Hozo et al. has some serious limitations and is always less satisfactory in practice because the estimator does not fully incorporate the sample size. As we explained in Section 'Estimating $\bar{X}$ and $S$ from $\mathcal{C}_1$', using $(b - a)/6$ for $n > 70$ in Hozo et al.'s adaptive estimation is untenable because the range $b - a$ tends to be infinity as $n$ approaches infinity if the distribution is not bounded, such as the normal and log-normal distributions. Our estimator replaces the adaptively selected thresholds ($\sqrt{12}, 4, 6$) with a unified quantity $2\Phi^{-1}((n - 0.375)/(n + 0.25))$, which can be quickly computed and obviously is more stable and adaptive. In addition, our method removes the non-negative

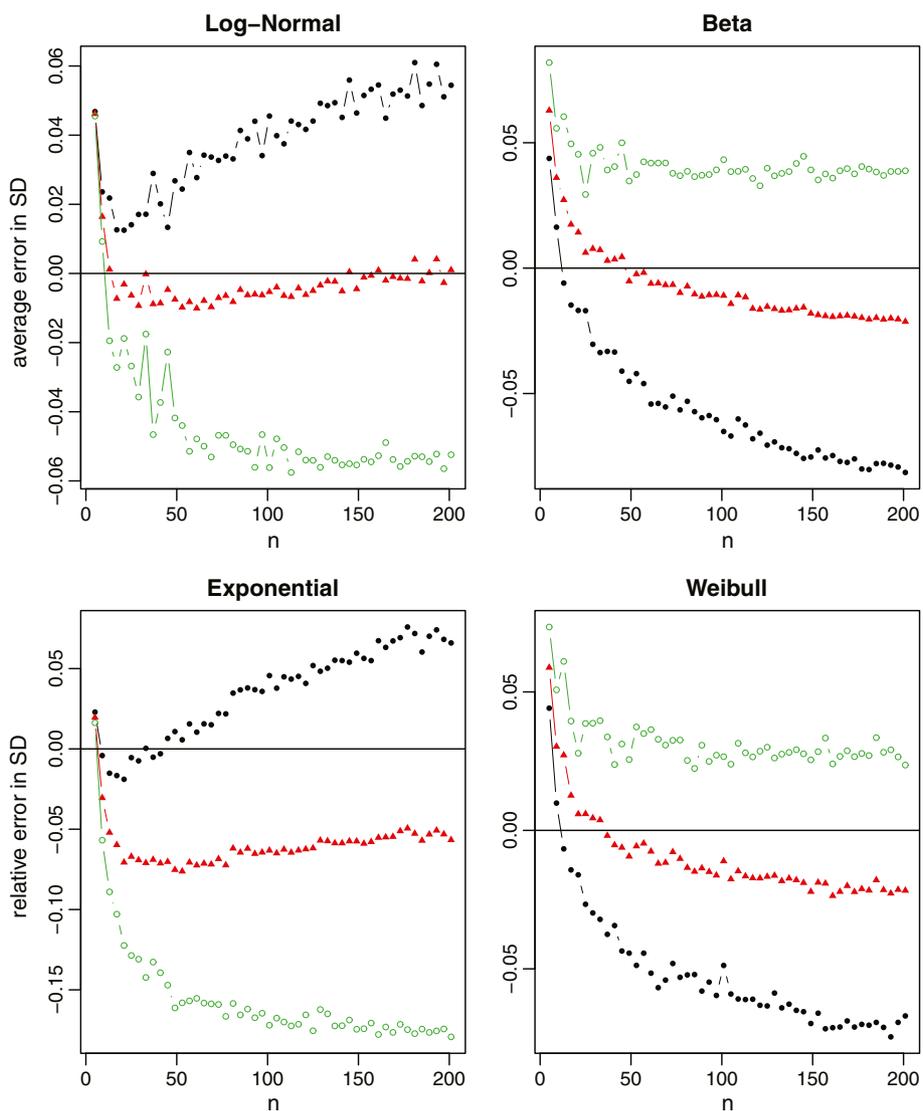

**Figure 6 Relative errors of the sample standard deviation estimation for non-normal data (log-normal, beta, exponential and Weibull),** where the black lines with solid circles represent the method under scenario $\mathcal{C}_1$, the red lines with solid triangles represent the method under scenario $\mathcal{C}_2$, and the green lines with empty circles represent the method under scenario $\mathcal{C}_3$.

**Table 3 Summary table for estimating $\bar{X}$ and $S$ under different scenarios**

|  | Scenario $\mathcal{C}_1$ | Scenario $\mathcal{C}_2$ | Scenario $\mathcal{C}_3$ |
|---|---|---|---|
| Hozo et al. (2005) | $\bar{X}$: Eq. (3) | – | – |
|  | $S$: Eq. (6) | – | – |
| Bland (2013) | – | $\bar{X}$: Eq. (10) | – |
|  | – | $S$: Eq. (11) | – |
| New methods | $\bar{X}$: Eq. (3) | $\bar{X}$: Eq. (10) | $\bar{X}$: Eq. (14) |
|  | $S$: Eq. (9) | $S$: Eq. (13) | $S$: Eq. (16) |

data assumption in Hozo et al.'s method and so is more applicable in practice.

Bland's method extended Hozo et al.'s method by using the additional information in the IQR. Since extra information is included, it is expected that Bland's estimators are superior to those in Hozo et al.'s method. However, the sample size is still not considered in Bland's method for the sample standard deviation, which again limits its capability in real-world cases. Our simulation studies show that Bland's estimator significantly overestimates the sample standard deviation when the sample size is large while seriously underestimating it when the sample size is small. Again, we incorporate the information of the sample size in the estimation of standard deviation via two unified quantities, $4\Phi^{-1}((n-0.375)/(n+0.25))$ and $4\Phi^{-1}((0.75n-0.125)/(n+0.25))$. With some extra but trivial computing costs, our method makes significant improvement over Bland's method when the IQR is available.

Moreover, we pay special attention to an overlooked scenario where the minimum and maximum values are not available. We show that the methodology following the ideas in Hozo et al.'s method and Bland's method will lead to unbounded estimators and is not feasible. On the contrary, we extend the ideas of our proposed methods in the other two scenarios and again construct a simple but still valid estimator. After that, we take a step forward to compare the estimators of the sample mean and standard deviation under all three scenarios. For simplicity, we have only considered three most commonly used scenarios, including $\mathcal{C}_1$, $\mathcal{C}_2$ and $\mathcal{C}_3$, in the current article. Our method, however, can be readily generalized to other scenarios, e.g., when only $\{a, q_1, q_3, b; n\}$ are known or when additional quantile information is given.

## Additional files

**Additional file 1: The plot of each of those distributions in the simulation studies.**

**Additional file 2: An Excel spread sheet including all formulas.**

**Competing interests**
The authors declare that they have no competing interests.

**Authors' contributions**
TT, XW, and JL conceived and designed the methods. TT and WW conducted the implementation and experiments. All authors were involved in the manuscript preparation. All authors read and approved the final manuscript.

**Acknowledgements**
The authors would like to thank the editor, the associate editor, and two reviewers for their helpful and constructive comments that greatly helped improving the final version of the article. X. Wan's research was supported by the Hong Kong RGC grant HKBU12202114 and the Hong Kong Baptist University grant FRG2/13-14/005. T.J. Tong's research was supported by the Hong Kong RGC grant HKBU202711 and the Hong Kong Baptist University grants FRG2/11-12/110, FRG1/13-14/018, and FRG2/13-14/062.

**Author details**
[1]Department of Computer Science, Hong Kong Baptist University, Kowloon Tong, Hong Kong. [2]Department of Statistics, Northwestern University, Evanston IL, USA. [3]Department of Mathematics, Hong Kong Baptist University, Kowloon Tong, Hong Kong.

**References**
1. Antman EM, Lau J, Kupelnick B, Mosteller F, Chalmers TC: **A comparison of results of meta-analyses of randomized control trials and recommendations of clinical experts: treatments for myocardial infarction.** *J Am Med Assoc* 1992, **268:**240–248.
2. Cipriani A, Geddes J: **Comparison of systematic and narrative reviews: the example of the atypical antipsychotics.** *Epidemiol Psichiatr Soc* 2003, **12:**146–153.
3. Hozo SP, Djulbegovic B, Hozo I: **Estimating the mean and variance from the median, range, and the size of a sample.** *BMC Med Res Methodol* 2005, **5:**13.
4. Triola M. F: *Elementary Statistics, 11th Ed*. Addison Wesley; 2009.
5. Hogg RV, Craig AT: *Introduction to Mathematical Statistics*. Maxwell: Macmillan Canada; 1995.
6. David HA, Nagaraja HN: *Order Statistics, 3rd Ed*. Wiley Series in Probability and Statistics; 2003.
7. Blom G: *Statistical Estimates and Transformed Beta Variables*. New York: John Wiley and Sons, Inc.; 1958.
8. Harter HL: **Expected values of normal order statistics.** *Biometrika* 1961, **48:**151–165.
9. Cramér H: *Mathematical Methods of Statistics*: Princeton University Press; 1999.
10. Bland M: **Estimating mean and standard deviation from the sample size, three quartiles, minimum, and maximum.** *International Journal of Statistics in Medical Research,* in press. 2014.
11. Liu T, Li G, Li L, Korantzopoulos P: **Association between c-reactive protein and recurrence of atrial fibrillation after successful electrical cardioversion: a meta-analysis.** *J Am Coll Cardiol* 2007, **49:**1642–1648.
12. Zhu A, Ge D, Zhang J, Teng Y, Yuan C, Huang M, Adcock IM, Barnes PJ, Yao X: **Sputum myeloperoxidase in chronic obstructive pulmonary disease.** *Eur J Med Res* 2014, **19:**12.
13. Higgins JPT, Green S: *Cochrane Handbook for Systematic Reviews of Interventions*: Wiley Online Library; 2008.